\documentclass[aps,pre,twocolumn,showpacs,superscriptaddress,floatfix]{revtex4-1}
\usepackage{graphicx}
\usepackage{dcolumn}
\usepackage{bm}
\usepackage{amssymb}
\usepackage{amsmath}
\usepackage{color}
\usepackage{hyperref}
\usepackage{subfigure}
\usepackage[usenames,dvipsnames]{xcolor}
\newcommand{\be}{\begin{equation}}
\newcommand{\ee}{\end{equation}}
\usepackage[normalem]{ulem}

\hypersetup{
     colorlinks=true,       
     linkcolor=blue,          
     citecolor=blue,        
     filecolor=magenta,      
     urlcolor=black         
}

\begin{document}

\title{Deformation-Driven Diffusion and Plastic Flow in Two-Dimensional Amorphous Granular Pillars}

\author{Wenbin Li}
\affiliation{Department of Materials Science and Engineering,
  Massachusetts Institute of Technology, Cambridge, Massachusetts
  02139, USA}

\author{Jennifer M. Rieser} \affiliation{Department of
  Physics and Astronomy, University of Pennsylvania, Philadelphia,
  Pennsylvania 19104, USA}

\author{Andrea J. Liu}
\affiliation{Department of Physics and Astronomy, University of
  Pennsylvania, Philadelphia, Pennsylvania 19104, USA}

\author{Douglas J. Durian} \email{djdurian@physics.upenn.edu}\affiliation{Department of Physics and Astronomy,
  University of Pennsylvania, Philadelphia, Pennsylvania 19104, USA}

\author{Ju Li} \email{liju@mit.edu} \affiliation{Department of Nuclear Science and
  Engineering, Massachusetts Institute of Technology, Cambridge,
  Massachusetts 02139, USA} \affiliation{Department of Materials
  Science and Engineering, Massachusetts Institute of Technology,
  Cambridge, Massachusetts 02139, USA}

\begin{abstract}
  We report a combined experimental and simulation study of
  deformation-induced diffusion in compacted two-dimensional amorphous
  granular pillars, in which thermal fluctuations play negligible
  role. The pillars, consisting of bidisperse cylindrical acetal
  plastic particles standing upright on a substrate, are deformed
  uniaxially and quasistatically by a rigid bar moving at a constant
  speed. The plastic flow and particle rearrangements in the pillars
  are characterized by computing the best-fit affine transformation
  strain and non-affine displacement associated with each particle
  between two stages of deformation. The non-affine displacement
  exhibits exponential crossover from ballistic to diffusive behavior
  with respect to the cumulative deviatoric strain, indicating that in
  athermal granular packings, the cumulative deviatoric strain plays
  the role of time in thermal systems and drives effective particle
  diffusion. We further study the size-dependent deformation of the
  granular pillars by simulation, and find that different-sized
  pillars follow self-similar shape evolution during deformation. In
  addition, the yield stress of the pillars increases linearly with
  pillar size. Formation of transient shear lines in the pillars
  during deformation becomes more evident as pillar size increases.
  The width of these elementary shear bands is about twice the
  diameter of a particle, and does not vary with pillar size.
\end{abstract}


\maketitle

\section{Introduction}

Disordered materials such as metallic glasses can exhibit highly
localized deformation and shear band formation\cite{Greer13, Falk11}.
Most experiments on these systems, however, use loading geometries in
which there are free boundaries and inhomogeneous strains, while
simulations have typically focused on systems with periodic boundary
conditions under homogeneously-applied shear strain.  To understand at
a microscopic level the effects of loading geometry on the macroscopic
mechanical response, it is useful to study a disordered system in
which individual particles can be imaged and tracked as they rearrange
under an applied load.  Here we introduce a granular packing--a
packing of discrete macroscopic particles for which thermal agitation
plays a negligible role \cite{Jaeger96, DeGennes99}--in a pillar
geometry commonly used for mechanical testing of metallic glasses.  We
combine experiment and simulation to study the response of these
two-dimensional (2D) pillars to athermal, quasistatic, uniaxial
compression.

One question of interest is how the mechanical response of the pillar
depends on pillar size. We find that the pillar shape evolves under
load in a self-similar fashion, so that the shape of the pillar at a
given strain is independent of system size.  We also find that as the
pillars deform, the strain rate localizes into transient lines of
slip, whose thickness of a few particle diameters is independent of
system size. Thus, the system is self-similar in shape at the
macroscopic scale, but, surprisingly, its yielding is not self-similar
at the microscopic scale.


A second question concerns the random motions of particles as they
rearrange under inhomogeneous loading conditions.  Because particles
jostle each other, they display diffusive behavior in homogeneously
sheared systems that are devoid of random thermal
fluctuations~\cite{Ono02}. Recently, crystal nucleation and growth
were observed {\it in situ} in mechanically fatigued metallic glasses
at low temperature \cite{Wang13}. Crystallization is typically thought
to require diffusion.  Therefore, it was suggested that the ``shear
transformation zones" (STZs) ~\cite{Falk11} should be generalized to
``shear diffusion transformation zones" (SDTZs)~\cite{Wang13}, to
reflect the contributions of random motions driven by loading, even
under inhomogeneous conditions.  Our amorphous granular pillar is an
athermal system as far as the macroscopic particles are concerned
(effective vibrational temperature $\approx$ 0), so our experiment and
simulations can examine how inhomogeneous loading affects particle
motion. We find that the idea of load-induced diffusion can be
generalized to inhomogeneous loading by replacing time with the
cumulative deviatoric strain, and the mean-squared displacement with
the mean-squared displacement of a particle relative to the best-fit
affine displacement of its neighborhood (\emph{i.e.} the mean-squared
non-affine displacement~\cite{Falk98}).  With this generalization, we
observe that the mean-squared non-affine particle displacement crosses
over from ballistic to diffusive behavior as a function of the
cumulative deviatoric strain.

The article is organized as follows. In section~\ref{sec:method}, we
describe the experimental and simulation setup, as well as the
simulation methodology, of 2D amorphous granular pillars under
uniaxial and quasistatic deformation. Section~\ref{sec:smallpillar}
describes the results of our combined experiments and simulations on
the deformation of a 2D granular pillar containing 1000 particles. In
section~\ref{sec:strain_diffusion}, we discuss the exponential
crossover of non-affine particle displacement from ballistic motion to
diffusion with respect to cumulative deviatoric
strain. Section~\ref{sec:sizeDependence} presents our simulation
results on the size-dependent deformation of large 2D granular
pillars.  Then we conclude the article in
section~\ref{sec:conclusion}.

\section{Methods  \label{sec:method}}

\begin{figure}[t!]
\begin{centering}
\includegraphics[width=0.35\textwidth]{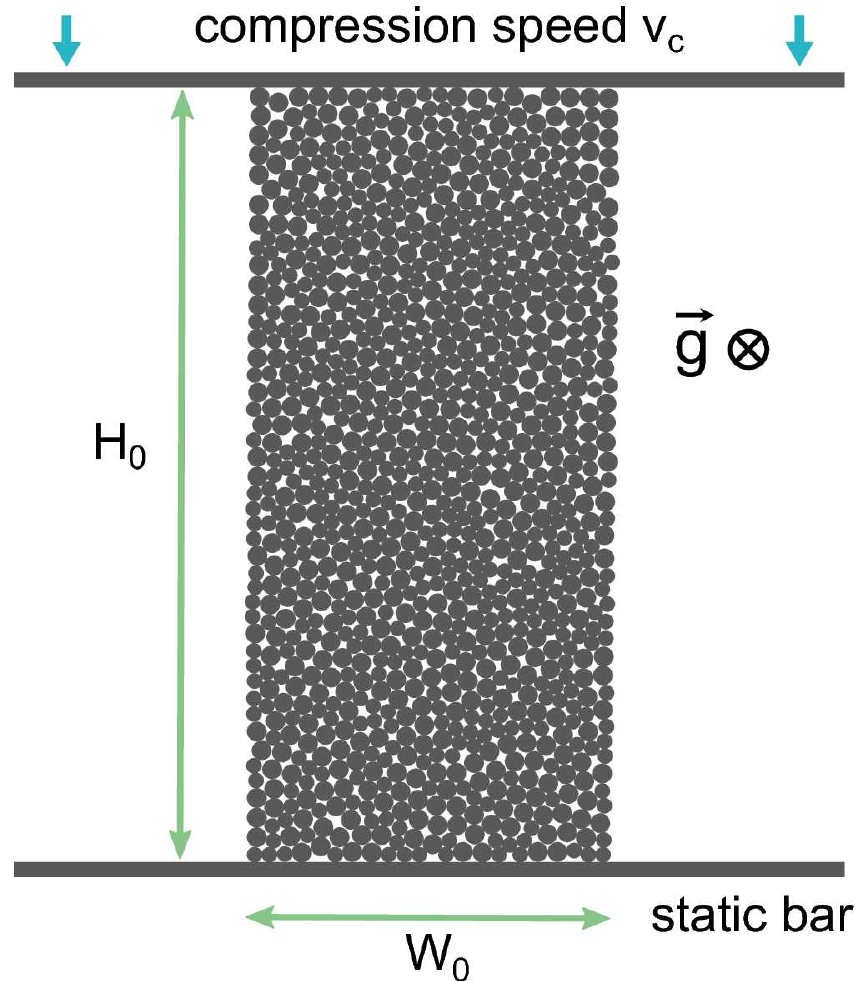}
\caption{Top-view of the experimental/simulation setup. A
  two-dimensional pillar of granular particles on a frictional
  substrate are deformed quasistatically and uniaxially by a rigid bar
  from one side. The direction of gravity is perpendicular to the
  substrate. The compacted, disordered granular packing consists of
  50-50 mixture of bidisperse cylindrical-shape grains.  The ratio of
  radius between large and small grains is 4:3. The aspect ratio of
  the pillar, defined as the initial height of the pillar ($H_0$)
  divided by the initial width ($W_0$), is 2:1. The pillar is confined
  between two rigid bars placed at the top and bottom end of the
  pillar respectively. The top bar deforms the pillar with a constant
  speed $v_c$ while the bottom bar is kept static.}
\label{fig:fig1}
\end{centering}
\end{figure}

The compacted 2D amorphous granular pillars in our study consist of
50-50 mixture of bidisperse cylindrical particles (grains) standing
upright on a substrate.  A top-view of the schematic setup is shown in
Fig.~\ref{fig:fig1}. The pillars have aspect ratio $H_0/W_0 \approx
2$, where $H_0$ and $W_0$ are the original height and width of the
pillars respectively. In our experiment, the cylindrical granular
particles are made of acetal plastic.  The diameter of the large
grains in the pillars, denoted by $D$, is $1/4$ inch (0.635 cm), while
for the small grains the diameter $d$ has the value of $3/16$ inch
(0.47625 cm). The ratio of diameter between large and small grains is
therefore $D/d = 4/3$. Both types of grains are 3/4 inch (1.905 cm)
tall. The masses for the large and small grains are 0.80 gram and 0.45
gram respectively. The pillars are confined between a pair of parallel
bars.  The bottom bar is static while the top bar deforms the pillars
uniaxially with a slow, constant speed $v_c = 1/300 $ inch per second
(0.0084667 cm/sec). The force sensors connected to the bars measure
the forces on the top and the bottom bars, and the trajectory of each
particle in a pillar is tracked by a high-speed camera mounted above
the pillar. The basic parameters in our simulation, including the size
and mass of the grains, as well as the velocity of the bars, are the
same as in the experiment. Further experimental details will be
described in an upcoming paper \cite{Rieser}.

\subsection{Packing Generation Protocol}
Properly prepared initial configurations are crucial for the study of
the mechanical properties of amorphous solids. In our experiment,
50-50 random mixture of bidisperse grains are compacted to form a
pillar with aspect ratio 2 to 1. To facilitate direct comparison
between experiment and simulation, for small-sized pillars (number of
grains in the pillar $N = 1000$), the simulation initial conditions
are taken from the experimental data, which was then relaxed in
simulation to avoid particle overlapping resulted from measurement
error. For large-sized pillars, which can only be studied by
simulation, we generate compacted, amorphous granular pillars through
computer simulation, using the protocol described below. The particle
area density in the simulation-generated pillar is controlled to be at
the onset of jamming transition \cite{Ohern03}. To generate the
initial conditions, we assign the following truncated Lennard-Jones
potential with purely repulsive interaction to the large (L) and small
(S) grains
\begin{equation}
  U_{\alpha\beta}(r) = \left\{
  \begin{array}{l l}
    \epsilon\left[\left(\sigma_{\alpha\beta}/r\right)^{12} - 2\left(\sigma_{\alpha\beta}/r\right)^{6} \right] & \quad \text{for $r < \sigma_{\alpha\beta}$}, \\
    -\epsilon & \quad \text{for $r \ge \sigma_{\alpha\beta}$},
  \end{array} \right.
\end{equation}
where the subscripts $\alpha$, $\beta$ denote L or S. The zero-force
cut-off distances $\sigma_{\alpha\beta}$ are chosen to be the sum of
radii of two particles in contact, namely $\sigma_{\text{LL}} = D$,
$\sigma_{\text{LS}} = 7D/8$, and $\sigma_{\text{SS}} = 3D/4$, where
$D$ is the diameter of a large grain. We note that this potential will
only be used to generate the initial conditions of the granular
packings, and is different from the particle interaction model we
describe later for the deformation of the granular pillars.

\begin{figure}[t!]
\begin{centering}
\includegraphics[width=0.45\textwidth]{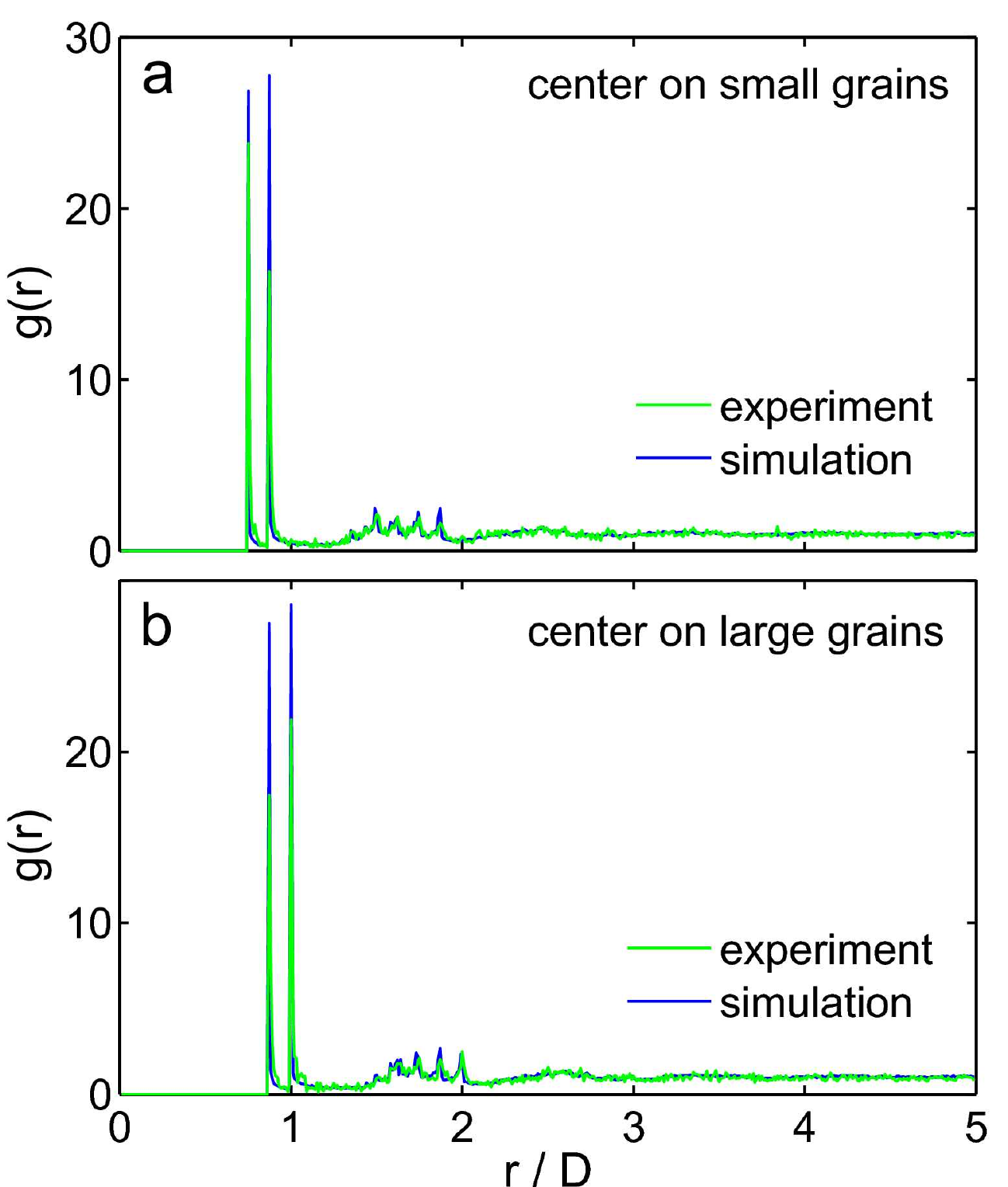}
\caption{\label{fig:rdf_coplot}Comparison of the radial distribution
  functions $g(r)$ for experiment-derived and simulation-generated
  initial conditions computed using (a) small grains as the
  central particles and (b) larger grains as the central particles are
  shown respectively. The distance $r$ is scaled by the
  diameter $D$ of the large particles.}
\end{centering}
\end{figure}

To create a disordered granular packing with 50-50 mixture of $N$
total number of large and small grains, a rectangular simulation box
with dimensions $\Lambda \times 2 \Lambda$ is initially created, where
the width of the box $\Lambda$ is chosen such that the initial
particle area density, $\rho = N/2\Lambda^2$, is slightly above the
particle overlapping threshold. We then randomly assign the positions
of the particle within the simulation box, and subsequently use
conjugate-gradient (CG) method to minimize the total potential energy
of the system. Periodic boundary conditions are applied during this
process. The particle positions are adjusted iteratively until the
relative change of energy per particle between two successive CG steps
is smaller than $10^{-12}$. When this stage is reached, the pressure
of the system is calculated using the following virial formula
\begin{equation}
p = -\frac{1}{2A}\sum_{i>j} r_{ij}\frac{dU}{dr_{ij}},
\end{equation}
where $A$ is the area of the simulation box, $r_{ij}$ is the distance
between particles $i$ and $j$. If the pressure is greater than zero,
both dimensions of the simulation box will be enlarged by a fraction
of $10^{-5}$, and the particles in the box will be mapped to the
corresponding new positions in the enlarged box via affine
transformation. CG energy minimization will then be carried out on the
new configuration. This iterative process stops when the calculated
pressure of the system at the end of a CG run becomes smaller than
$10^{-10} \epsilon/D^2 $. The final configuration will be taken as the
initial conditions of close-packed 2D amorphous granular
assembly. Vacuum space is then added on the lateral sides of
simulation box to create a pillar with 2:1 aspect ratio. Calculation
of radial distribution functions for different-sized pillars indicates
that the structure of the amorphous assemblies generated following the
above procedures does not show noticeable size dependence.  Comparison
of the radial distribution functions computed for the experimental and
simulation-generated initial conditions is shown in
Fig.~\ref{fig:rdf_coplot}.

\subsection{Simulation Methodology}
We use the method of MD to simulate the quasistatic deformation of the
2D granular pillars. The simulation force model includes three
components: the grain-grain interaction, the grain-bar interaction and
the grain-substrate interaction. Each of these forces will be
described in the following.

\subsubsection{Grain-Grain Interaction}

\begin{figure}[t!]
\begin{centering}
\includegraphics[width=0.45\textwidth]{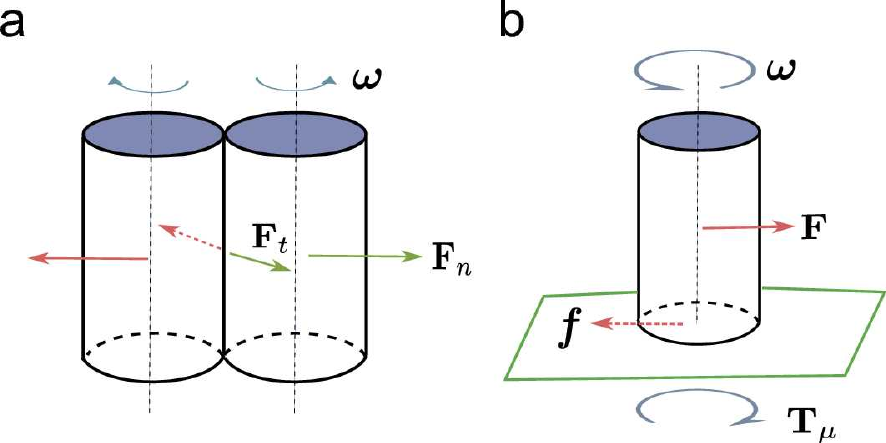}
\caption{\label{fig:fig3} (a) Illustrations of grain-grain interaction
  in the granular pillar. The contact force between two grains
  consists of normal repulsive contact force $\mathbf{F}_n$ and
  tangential shear contact force $\mathbf{F}_t$. (b) Illustration of
  grain-substrate interaction. If the velocity of a grain $i$ is
  non-zero, or the vector sum of the forces on the grain due to other
  grains and the bars is non-zero, the substrate will exert a
  frictional force $\mathbf{f}$ on the grain, the maximum value of
  which is $m_ig\mu$, where $m_i$ is the mass of the particle, $g$ is
  the gravity acceleration constant and $\mu$ denotes the friction
  coefficient between the grain and the substrate. Likewise, if the
  angular velocity of the grain is non-zero or the torque on the grain
  due to other interactions is non-zero, the substrate will induce a
  frictional torque whose maximum magnitude is $|\mathbf{T}_{\mu,i}| =
  \frac{2}{3}m_ig\mu R_i $, where $R_i$ is the radius of the
  particle.}
\end{centering}
\end{figure}


As illustrated in Fig.~\ref{fig:fig3}a, the interaction between two
grains includes normal and tangential contact force, which are denoted
by $\mathbf{F}_n$ and $\mathbf{F}_t$ respectively. Two grains
experience a repulsive normal contact force if the distance between
the particle centers is smaller than the sum of their radii. For two
smooth, elastic cylindrical particles with parallel axes, the normal
contact force as determined by the Hertzian theory of contact
mechanics is proportional to the indentation depth between the two
particles \cite{Johnson85}. For our granular particles, denote by
$\mathbf{r}_i$ and $\mathbf{r}_j$ the positions of particles $i$ and
$j$, and denote by $\mathbf{r}_{ij} = \mathbf{r}_i - \mathbf{r}_j$ the
distance vector between the two particles, the indentation depth
$\delta_{ij}$ is calculated as
\begin{equation}
\delta_{ij} = R_i + R_j - r_{ij},
\end{equation}
where $r_{ij} = |\mathbf{r}_{ij}|$. $R_i$ and $R_j$ are the particle
radius of $i$ and $j$ respectively. $\delta_{ij}$ will be zero if the
two particles are not in contact. The normal contact force acting on
the particle $i$ by particle $j$ is then given by
\begin{equation}
\mathbf{F}_{n_{ij}} = k_n \delta_{ij}\mathbf{n}_{ij},
\end{equation}
where $\mathbf{n}_{ij} = \mathbf{r}_{ij}/r_{ij}$, and $k_n$ is the
normal contact stiffness. The corresponding normal contact force on
particle $j$ is given by Newton's third law, namely,
$\mathbf{F}_{n_{ji}} = - \mathbf{F}_{n_{ij}}$.  In Hertzian theory of
contact mechanics \cite{Johnson85}, the constant $k_n$ for two
cylinders in contact can be calculated as
\begin{equation}
  k_n = \frac{\pi}{4}E^*l,
\label{eq:normal_modulus}
\end{equation}
where $l$ is the height of the cylinders. $E^*$ is the normalized
contact elastic modulus, which is computed from the respective elastic
modulus of the two cylinders, $E_1$ and $E_2$, and their Poisson's
ratios, $\nu_1$ and $\nu_2$:
\begin{equation}
  \frac{1}{E^*} = \frac{1-\nu_1^2}{E_1} + \frac{1-\nu_2^2}{E_2}.
\label{eq:effective_stiffness}
\end{equation}

The existence of a friction force between two particles in contact is
a characteristic feature of granular materials. Appropriate modeling
of contact friction is crucial to the study of granular dynamics. The
tangential frictional force between two grains in contact can be very
complicated in reality \cite{Silbert01}. We adopt the
history-dependent shear contact model initially developed by Cundall
and Strack \cite{Cundall79}. This well-tested model has been used by
many others to model the dynamics of granular assemblies
\cite{Silbert01, Silbert02, Landry03, Brewster05, Zhang05,
  Rycroft06Grest, Rycroft06Bazant, Kamrin07, Rycroft09Orpe,
  Rycroft09Kamrin}. The essence of this model is to keep track of the
elastic shear displacement of two particles throughout the lifetime of
their contact, and applying the Coulomb elastic yield criterion when
the displacement reaches a critical value. Our implementation of the
Cundall-Strack model follows Silbert \textit{et al.} \cite{Silbert01}.
Specifically, the tangential contact force between particle $i$ and
$j$ is calculated as:
\begin{equation}
\mathbf{F}_{t_{ij}} = -k_t\mathbf{u}_{t_{ij}},
\end{equation}
where the shear displacement $\mathbf{u}_{t_{ij}}$ is obtained by
integrating the tangential relative velocities of the two particles
during the lifetime of their contact \cite{Silbert01}. Here $k_t$ is
the tangential contact elastic modulus. It is taken to be proportional
to the normal contact stiffness $k_n$. Following Silbert \textit{et
  al.}, we choose $k_t = \frac{2}{7}k_n$. Previous studies showed that
the dynamics of system is relatively insensitive to this parameter
\cite{Silbert01}, which is confirmed by our own simulation.

To model the elastic yield of shear contact, the magnitude of
$\mathbf{u}_{t_{ij}}$ is truncated to satisfy the Coulomb yield
criterion $|\mathbf{F}_{t_{ij}}| \leq |\mu_g \mathbf{F}_{n_{ij}}|$,
where $\mu_g$ is the friction coefficient between the grains.

The tangential contact force will induce torques on the two grains in
contact, as given by
\begin{equation}
\mathbf{T}_{ij} = -\frac{1}{2} \mathbf{r}_{ij}\times \mathbf{F}_{t_{ij}}.
\end{equation}
Here $\mathbf{T}_{ij}$ is the torque exerted by grain $j$ on grain $i$
due to the tangential contact force $\mathbf{F}_{t_{ij}}$.

\subsubsection{Grain-Bar Interaction}
The grain-bar interaction is modeled in a similar way to the
grain-grain interaction. The bar is essentially treated as a rigid
grain with infinitely large radius.  When a grain comes in contact
with a bar, the grain can experience normal and shear contact force
induced by the bar, and the shear contact force is also calculated by
tracking the elastic shear displacement between the grain and the
bar. The motion of the moving bar is not affected by the grains. The
static bar at the bottom side of the pillar is always static, while
the top bar deforms the pillar at a constant speed $v_c$.  Compared to
grain-grain interaction, the interaction parameters between the grains
and the bar is slightly modified. Since the bars are modeled as rigid
bodies that cannot be elastically deformed, it means that the elastic
modulus of the bars is considered to be infinite. Consequently, the
effective interaction modulus $E^*$ between the bars and the grains,
based on Eq.~\ref{eq:effective_stiffness}, is twice as large as that
between the grains. Therefore, from Eq.~\ref{eq:normal_modulus}, the
normal interaction stiffness between the bars and the grains is twice
as large as that between the grains, i.e., $k_n(\text{grain-bar}) =
2k_n(\text{grain-grain})$. Since the shear modulus of contact $k_t$ is
proportional to $k_n$, we have $k_t(\text{grain-bar}) =
2k_t(\text{grain-grain})$ as well.

\subsubsection{Grain-Substrate Interaction}
The effect of the substrate on the grains is determined after all the
forces and torques on each grain due to other grains and bars have
been determined.  The substrate can induce both frictional force and
torque on the grains, as illustrated in Fig.~\ref{fig:fig3}b. If a
grain is initially static, unless the magnitude of total force due to
other grains/bars is larger than the maximum frictional force that can
be exerted by the substrate $|\mathbf{f}_i| = m_ig\mu$, the substrate
frictional force will cancel out other forces on the grain and the
particle will continue to have zero velocity. Here $m_i$ is the mass
of the grain $i$, $g$ is the gravitational acceleration and $\mu$
denotes the frictional coefficient between the grains and the
substrate. In another case, if the velocity of the grain is non-zero,
the substrate will induce a frictional force opposite to the direction
of particle motion, with magnitude $|\mathbf{f}_i| = m_ig\mu$.  A
similar algorithm applies to the rotational motion of a particle. An
initially static grain will not start to rotate unless the torque due
to other interactions surpasses the maximum substrate-induced
frictional torque $|\mathbf{T}_{{\mu}_i}| = \frac{2}{3}m_ig\mu R_i$,
where $R_i$ is the radius of the cylindrical-shape particle. The
prefactor $\frac{2}{3}$ is based on the assumption that frictional
force is evenly distributed on the circular contact interface between
a cylindrical-shape grain and the substrate. If the angular velocity
of the grain is non-zero, a frictional torque
\begin{equation}
\mathbf{T}_{{\mu},i} = -\frac{2}{3}m_ig\mu R_i \hat{\boldsymbol \omega_i},
\end{equation}
will slow down the rotational motion of the particles, where
$\hat{\boldsymbol \omega_i} = {\boldsymbol \omega_i}/|{\boldsymbol
  \omega_i}|$ and $\boldsymbol \omega_i$ denotes the angular velocity
of particle $i$.

\subsubsection{Equations of Motion}
After all the forces and torques on an individual grain are
determined, they are summed up and the velocities and angular
velocities of the grains are then updated according to Newtonian
equations of motion:
\begin{equation}
  m_i \frac{d^2\mathbf{r}_i}{dt^2} =
  \mathbf{F}_i, \qquad I_i \frac{d\boldsymbol\omega_i}{dt} =
  \mathbf{T}_i,
\label{eq:EOM}
\end{equation}
where $\mathbf{F}_i$ and $\mathbf{T}_i$ are the total force and torque
on the particle $i$ respectively. $I_i = \frac{1}{2}m_i R_i^2$ is the
moment of inertia for grain $i$. The standard velocity Verlet
integrator is used to update the positions and velocities of the
particles, while a finite difference method is used to integrate the
first-order differential equation for the angular velocities.

There is a subtle numerical issue that must be addressed when modeling
velocity and angular velocity changes of the particles in the presence
of the damping effects of a frictional substrate. In numerical
integration of equation of motion, time is discretized into small
timesteps with each timestep being a small increment $\delta t$. To
complete the simulation within a reasonable time frame, $\delta t$
cannot be too small, which means that the changes of velocity and
angular velocity of the grains due to the substrate induced force and
torque within a timestep are not infinitesimal. Hence, the motion of
particles might not be able to be brought to a halt by the substrate
$-$ the velocity and angular velocity of the particles could oscillate
around the zero. Consider, for example, a stand-alone cylindrical
grain with initial velocity $\mathbf{v}_i$ and angular velocity
$\boldsymbol \omega_i$. Without other interactions, the substrate will
induce friction $|\mathbf{f}_i| = m_ig\mu$ and frictional torque
$|\mathbf{T}_{\mu, i}| = 2m_ig\mu/3$ on the grain, which slows down
the translational and rotational motion of the grain
respectively. According to the equations of motion in
Eq.~\ref{eq:EOM}, the translational and rotational acceleration will
be $a_v = g\mu$ and $a_{\omega} = 4g\mu/(3R_i)$, with $R_i$ being the
radius of particle $i$. Hence, within a timestep $\delta t$, the
change of velocity or angular velocity is a finite number: $\delta v =
g\mu \delta t$, $\delta \omega = 4g\mu \delta t/(3R_i)$. If the
velocity or angular velocity have been damped to values below these
two numbers, they cannot be damped further but instead oscillate
around zero, which is clearly a numerical artifact. To work around
this issue, we introduce two small parameters
\begin{equation}
\xi_v = g\mu \delta t, \quad \xi_{{\omega}_i} = \frac{4g\mu}{3R_i}\delta t,
\end{equation}
such that when $|\mathbf{v}_i| < \xi_v$ and $|\sum_j\mathbf{F}_{ij} +
\mathbf{F}_i^{\text{bar}}| \le m_i g\mu$ are both satisfied, the
velocity and total force on the particle will be set to zero. Here
$\mathbf{F}_{ij}$ is the force of particle $j$ on particle $i$, and
$\mathbf{F}_i^{\text{bar}}$ is the force of the bars on particle $i$.
Similarly, for the rotational motion, if $|\boldsymbol \omega_i| <
\xi_{{\omega}_i}$ and $|\sum_j\mathbf{T}_{ij} +
\mathbf{T}_i^{\text{bar}}| \le \frac{2}{3}m_i g \mu R_i$, the angular
velocity and total torque of the particle is set to zero.

\subsection{Choice of Simulation Model Parameters}

The independent parameters in the interaction model of our simulation
include the grain-grain stiffness $k_n$, grain-grain friction
coefficient $\mu_g$, grain-substrate friction coefficient $\mu$, and
the timestep for integration of equations of motion $\delta t$. Among
these parameters, $\mu$ has been experimentally measured to be around
0.23. Hence $\mu = 0.23$ will be adopted in our simulations. The
grain-grain friction coefficient $\mu_g$ is unknown. We have carried
out simulation using multiple values of $\mu_g$, and the results
indicate that choosing $\mu_g = 0.2$ can achieve good match between
the experiment and simulation. Due to the quasistatic nature of
deformation by the moving bar on the pillars, the increment of force
on a grain by the bar within one timestep $\delta t$ must be much
smaller than the maximum static friction by the substrate on a grain,
namely
\begin{equation}
2k_nv_c\delta t \ll m_ig\mu,
\label{eq:stiffness_bound}
\end{equation}
where $v_c$ is the speed of the top moving bar.  Hence, the smaller
the value of $\delta t$, the higher the value of $k_n$ that can be
adopted in simulation. While there is no physical reason for a lower
bound of $\delta t$, smaller $\delta t$ results in an increased time
span to complete simulation. Realistic consideration leads to our
choice of $\delta t = 10^{-5}$ second. The upper bound of allowed
$k_n$ calculated from Eq.~\ref{eq:stiffness_bound} is considered to be
smaller than the real contact stiffness of two particles in
experiment. For this reason, we have systematically studied the
influence of $k_n$ on the simulation results in a small-sized pillar
containing 1000 grains. The relatively small sized pillar allows us to
use $\delta t = 10^{-6}$ second and thus access a wider range of
$k_n$, from $k_n$ = 1 N/mm to $k_n$ = 100 N/mm. The results indicate
that the statistical behaviors of deformation dynamics, such as flow
stress and particle-level deformation characteristics, are not
significantly influenced by the value of the $k_n$. We therefore
choose $k_n = 10$ N/mm and $\delta t = 10^{-5}$ in our simulation.

The results of our study will be expressed in terms of several
characteristic units.  Length will be expressed in the diameter of the
large grains $D$ or the radius $R=D/2$.  The unit of velocity will be
the bar speed $v_c$ and the unit of time will be $R/v_c$, which is the
time it takes for top bar to move over a distance equal to $R$.  The
units for force and stress will be $mg\mu$, $mg\mu/D$ respectively,
where for convenience, we will use the symbol $m$ to denote the mass
of a large grain. $mg\mu$ is thus the minimum force to induce the
translational motion of a stand-alone large grain and $mg\mu/D$ is the
corresponding averaged stress of the bar on the grain.

\section{Combined Experiment and Simulation on Deformation of
  Small-Sized Pillars \label{sec:smallpillar}}

Deformation of an $N = 1000$ pillar has been studied by both
experiment and simulation. The experimental initial particle
arrangement in the pillar is the same as those depicted in
Fig.~\ref{fig:fig1}. To facilitate comparison between experiment and
simulation, our parallel simulation of pillar deformation uses the
experimentally measured initial conditions, which were further relaxed
in simulation to avoid particle overlapping resulted from measurement
error.  When the pillar is deformed by the moving bar, the strain of
deformation $\varepsilon$ is defined as the change of pillar height
$\Delta H$ divided by the original height of the pillar $H_0$, namely,
$\varepsilon \equiv \Delta H/H_0$. The deformation stress $\sigma$ is
calculated as the normal force on the top moving bar
$F_{\mathrm{bar}}$ divided by the maximum width of the pillar near the
top edge $W$, namely $\sigma \equiv F_{\mathrm{bar}}/W$.

\begin{figure}[t!]
\begin{centering}
\includegraphics[width=0.45\textwidth]{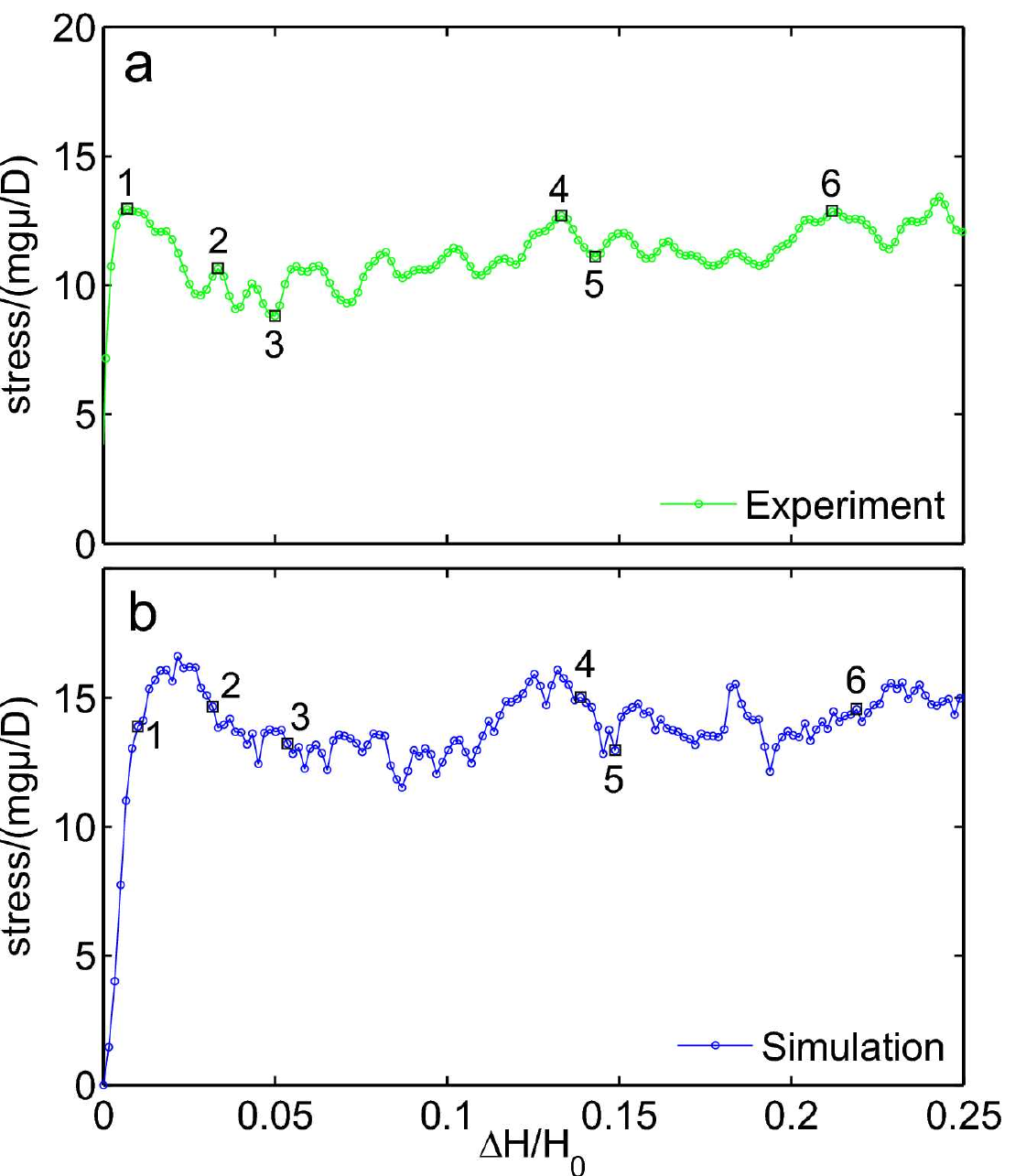}
\caption{\label{fig:forceCurve}Comparison between the (a) experimental
  and (b) simulation stress-strain curves for the deformation of a $N
  = 1000$ granular pillar. The compressing stress is measured in units
  of $mg\mu/D$, while the strain is computed as the change of pillar
  height ($\Delta H$) divided by the original height of the pillar
  $H_0$. The special labels (1-6) indicate the stress strain values at
  which deformation characteristics in the pillar will be compared
  side to side between experiment and simulation.}
\end{centering}
\end{figure}

Fig.~\ref{fig:forceCurve} shows the experimental and simulation
stress-strain curve of the $N = 1000$ pillar.  The measured stress
shows yielding behavior when the deformation strain exceeds a very
small value $\varepsilon_y$. From our simulation, we find that the
yield strain $\varepsilon_y$ in general becomes smaller as the
grain-grain stiffness $k_n$ or the packing density of the pillar is
increased. The yield stress $\sigma_y$ however shows little dependence
on $k_n$.  The parameter that affects $\sigma_y$ most was found to be
the grain-grain friction coefficient $\mu_g$. In the range of $\mu_g$
we have studied ($\mu_g$ from 0 to 0.3), $\sigma_y$ increases
monotonically with the increase of $\mu_g$. The simulation results
presented in this paper use $\mu_g = 0.2$, which was found to achieve
overall good match between the experiment and simulation.

\begin{figure*}[t!]
\begin{centering}
\includegraphics[width=0.9\textwidth]{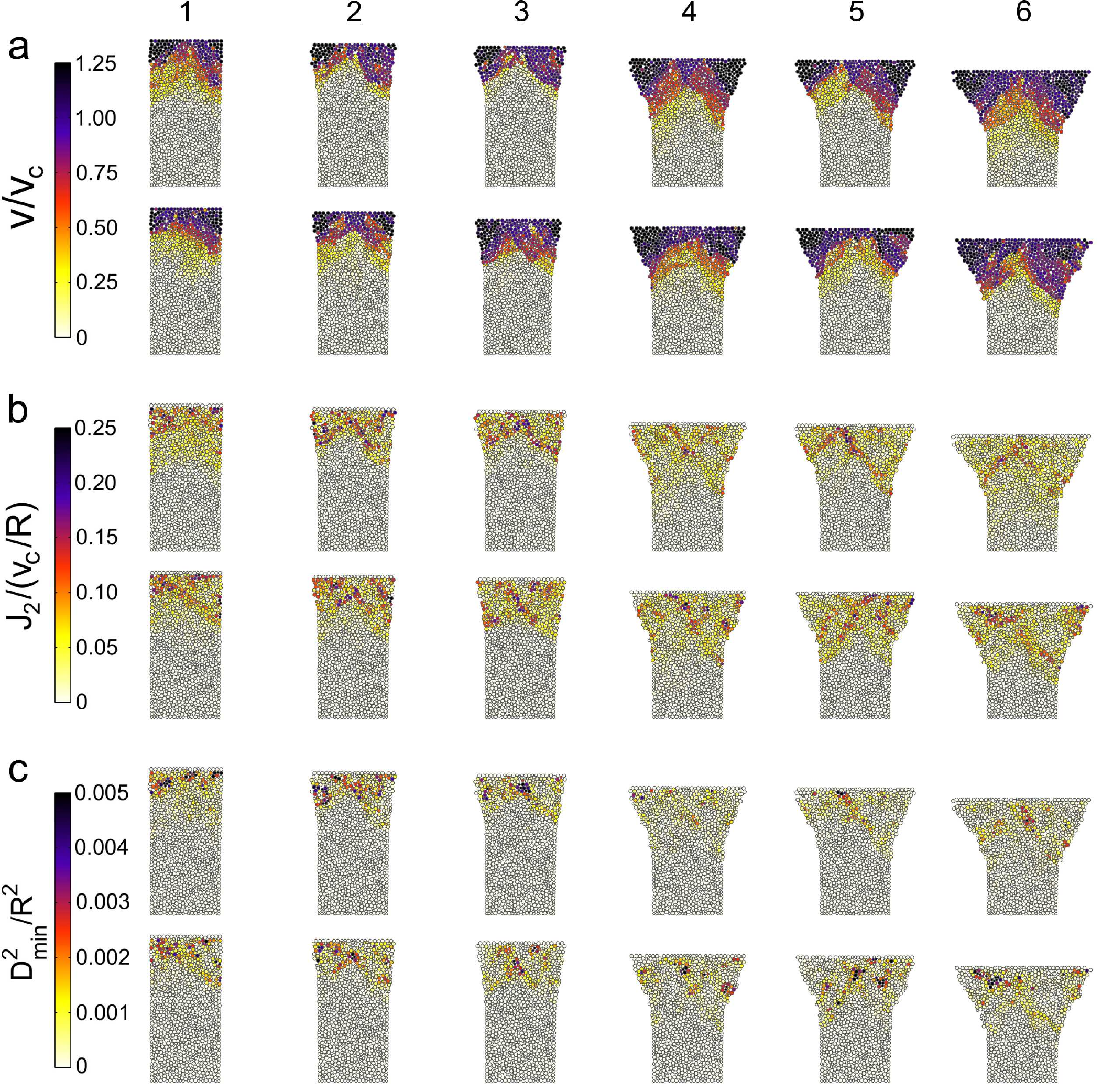}
\caption{\label{fig:colorCodedQuantities} Comparison between
  experiment and simulation of the particle velocity $v$, deviatoric
  strain rate $J_2$ and non-affine displacement $D^2_{\mathrm{min}}$
  during deformation of a $N = 1000$ granular pillar. The six stages
  of deformation (1-6) correspond to the stress and strain values
  labeled in Fig.~\ref{fig:forceCurve}. Within each subplot (a), (b)
  and (c), the top panel corresponds to the experimental result, while
  the bottom panel corresponds to the simulation result.  (a)
  Velocities of the particles in the pillar. The magnitude of the
  displacement of a particle from the current position after time
  interval $\Delta t = (2/15)R/v_c$ is divided by $\Delta t$ to obtain
  the average velocity across the time interval. (b) Deviatoric strain
  rate $J_2$ for each particle. $J_2$ is computed by comparing the
  current configuration of a particle and its neighbors with the
  configuration after $\Delta t$, using neighbor sampling distance
  $R_c = 1.25D$. $J_2$ is measured in the unit of $v_c/R$. (c)
  Non-affine displacement $D^2_{\mathrm{min}}$ for each particle in
  the pillar.  The procedures for calculating $D^2_{\mathrm{min}}$ are
  discussed in the main text.}
\end{centering}
\end{figure*}

In Fig.~\ref{fig:forceCurve}, we label several stress/strain values
and calibrate the corresponding particle-level structural changes in
the pillar. The experimental and simulation results are then compared
side-to-side in Fig.~\ref{fig:colorCodedQuantities}.
Fig.~\ref{fig:colorCodedQuantities}a shows the mean particle velocity
field in the pillar at six different stages of deformation. The mean
velocity of a particle $i$, denoted by $v_i(t, \Delta t)$, is
calculated as the average displacement magnitude of the particle from
current time $t$ to a later time $t+\Delta t$,
\begin{equation}
  v_i(t, \Delta t) = |\mathbf{r}_i(t+\Delta t) - \mathbf{r}_i(t)|/\Delta t,
\label{eq:meanVelocity}
\end{equation}
where the value of time interval $\Delta t$ is chosen to be $2/15$
$R/v_c$ for the present purpose. $v_i(t, \Delta t)$ contains
information of the absolute amount of displacement of the particle $i$
within $\Delta t$. As shown in Fig.~\ref{fig:colorCodedQuantities}a,
the mean velocities of the particles near the moving bar are close to
$v_{c}$, which is expected as the pillar is deformed quasistatically
by the bar. The mean velocity of a particle in general becomes smaller
as the particle is further away from the moving bar. At the early
stages of deformation, particles at the bottom part of the pillar have
not moved and therefore have zero values of $v$.  A sharp boundary
between the moving and non-moving regions of the pillar often forms
along the the direction that is roughly 45~degree to the direction of
uniaxial deformation.

\begin{figure*}[t!]
\begin{centering}
\includegraphics[width=0.95\textwidth]{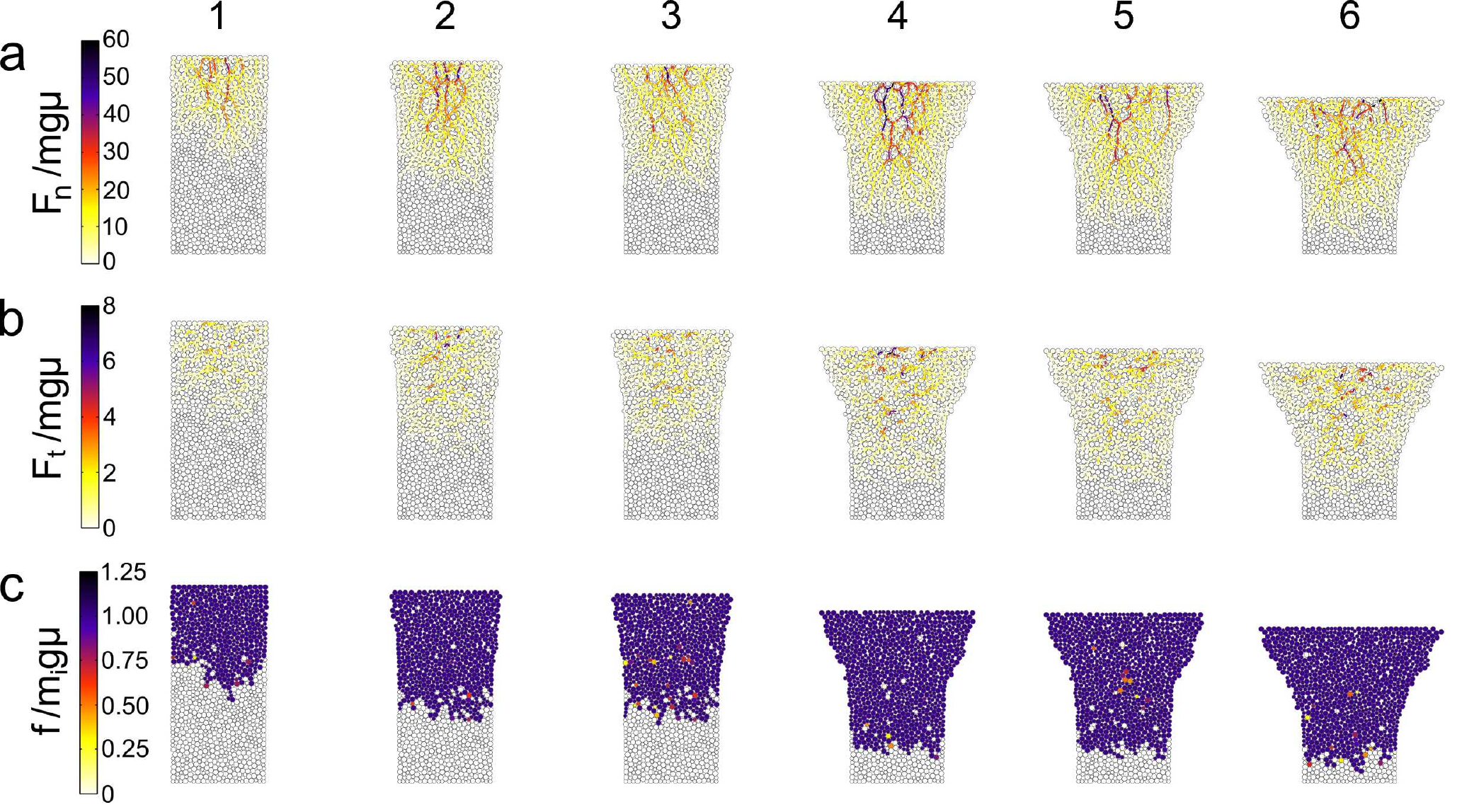}
\caption{\label{fig:simulationforce} Forces in the granular pillar
  during deformation as obtained from simulation. The special labels
  (1-6) correspond to the deformation stages labeled in
  Fig.~\ref{fig:forceCurve}b. (a) Grain-grain normal force $F_n$; (b)
  grain-grain tangential force $F_t$ and (c) grain-substrate friction
  force $f$. The normal and tangential forces are measured in the unit
  of $mg\mu$, which is the largest possible value of substrate induced
  friction on a large grain. The substrate friction forces are
  measured in the unit of $m_ig\mu$.}
\end{centering}
\end{figure*}

In the simulations we have access to detailed
information on the inter-particle interactions. In
Fig.~\ref{fig:simulationforce} we plot the grain-grain normal force
$F_n$, tangential force $F_t$ and substrate-induced force frictional
force $f$ on the particles at six stages of deformation corresponding
to the special labels in Fig.~\ref{fig:forceCurve}.  Comparing
Fig.~\ref{fig:simulationforce}a with Fig.~\ref{fig:simulationforce}b
and Fig.~\ref{fig:simulationforce}c, we find that $F_n$ is in general
much larger than $F_t$, which is further larger than $f$, namely $F_n
\gg F_t \gg f$.  In particular, Fig.~\ref{fig:simulationforce}a shows
that particles with large $F_n$ are connected with force chains. The
magnitude of forces in these force chains is higher for particles
residing in the interior the pillar.  This indicates that the stress
in the pillar is rather inhomogeneous, with larger stresses in the interior region of the
pillars than close to the surface.

\begin{figure}[t!]
\begin{centering}
\includegraphics[width=0.45\textwidth]{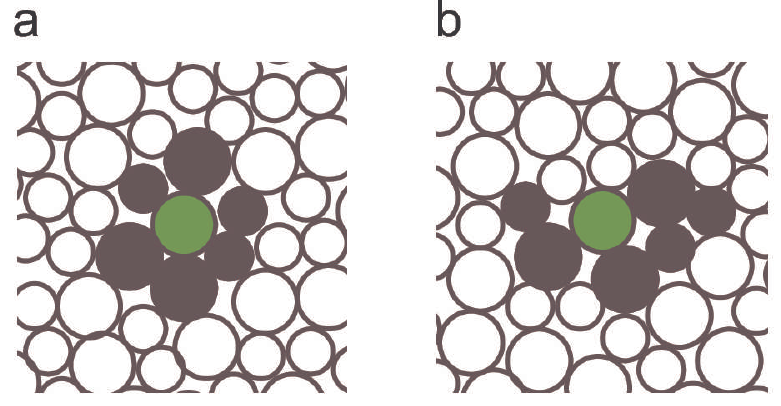}
\caption{\label{fig:fig7} (a) Illustration of a particle (colored in
  green) and its neighbors (colored in black) within a cut-off
  distance $R_c = 1.25 D$ at an initial reference configuration.  (b)
  The same set of particles at a later stage of deformation. We seek
  to find the best-affine transformation matrix $\mathbf{J}$ that maps
  the coordinates of the particles illustrated in (a) to those in
  (b). This optimization procedure also gives the non-affine
  displacement $D^2_{\mathrm{min}}$ associated with the central
  (green) particle, and the deviatoric strain $\eta^s$ in the
  neighborhood, as discussed in the main text.}
\end{centering}
\end{figure}

We further look at the rearrangement of particles in the pillar by
defining a neighbor sampling distance $R_c$, and calculate the affine
transformation strains and non-affine displacements of the particles
with respect to their neighbors within $R_c$. The value of $R_c$ is
chosen to be $1.25D$, which roughly corresponds to the average first
nearest-neighbor distance of the particles in the pillar, as can be
seen from the computed radial distribution functions in
Fig.~\ref{fig:rdf_coplot}. A particle $j$ is considered to be the
neighbor of a particle $i$ if their distance is smaller than $R_c$,
which is illustrated in Fig.~\ref{fig:fig7}.  The configurations of
the particle $i$ and its neighbors at a given time $t$ and a
subsequent time $t+\Delta t$ will then be used to compute the best-fit
local affine transformation matrix $\mathbf{J}$ and the non-affine
displacement $D^2_{\mathrm{min}}$ associated with particle $i$, using
the method introduced by Falk and Langer \cite{Falk98,
  Shimizu07}. Specifically, $D^2_{\mathrm{min}, i}$ is obtained by
calculating the best affine transformation matrix $\mathbf{J}_i$ that
minimizes the error of deformation mapping:
\begin{equation}
  D^2_{\mathrm{min},i}(t, \Delta t) = \frac{1}{N_i}\underset{\mathbf{J}_i}{\mathrm{min}} \sum_{j\in N_i}
  \left[\mathbf{r}_{ji}(t+\Delta t)- \mathbf{J}_i\mathbf{r}_{ji}(t)\right]^2,
\end{equation}
where $\mathbf{r}_{ji}(t) = \mathbf{r}_j(t) - \mathbf{r}_i(t)$ is the
distance vector between particles $j$ and $i$ at time
$t$. $\mathbf{r}_{ji}(t+\Delta t)$ is the distance vector at a later
time $t+\Delta t$. The summation is over the neighbors of particle $i$
at time $t$, whose total number is given by $N_i$. The best-fit affine
transformation matrix $\mathbf{J}_i(t,\Delta t)$ is usually
non-symmetric due to the presence of rotational component. A symmetric
Lagrangian strain matrix $\boldsymbol \eta_i$ can be calculated from
$\mathbf{J}_i$ as
\begin{equation}
\boldsymbol \eta_i = \frac{1}{2}\left(\mathbf{J}_i^T\mathbf{J}_i - \mathbf{I}\right),
\end{equation}
where $\mathbf{I}$ is an identity matrix. The
hydrostatic invariant is then computed from $\boldsymbol \eta_i$ as
\begin{equation}
\eta_i^{m} = \frac{1}{2}\mathrm{Tr}\boldsymbol \eta_i.
\end{equation}
The shear (deviatoric) invariant is then given by
\begin{equation}
  \eta_i^s = \sqrt{\frac{1}{2}\mathrm{Tr}\left(\boldsymbol{\eta}_i - \eta_i^{m}\mathbf{I}\right)^2}.
\end{equation}
Hereafter we will refer to $\eta_i^s(t, \Delta t)$ as the
deviatoric strain associated with the particle $i$ from $t$ to
$t+\Delta t$. The deviatoric strain rate, denoted by $J_2$, is the
normalization of $\eta_i^s(t, \Delta t)$ with respect to $\Delta t$:
\begin{equation}
  J_2(t, \Delta t) = \eta^s(t, \Delta t)/\Delta t
\label{eq:J2}
\end{equation}

Fig.~\ref{fig:colorCodedQuantities}b-c shows the computed deviatoric
strain rate $J_2$ and $D^2_{\mathrm{min}}$ for each particle in the
pillar at six different stages of deformation, where the experimental
and simulation results are compared side to side. $J_2(t, \Delta t)$
and $D^2_{\mathrm{min}}(t, \Delta t)$ are computed using $\Delta t
=(2/15) R/v_c$, which is the same as the value of $\Delta t$ used for
computing the mean velocities of the particles in
Fig.~\ref{fig:colorCodedQuantities}a. Comparing
Fig.~\ref{fig:colorCodedQuantities}b with
Fig.~\ref{fig:colorCodedQuantities}a, it can be seen that large values
of deviatoric strain rate occur at places where the gradient of mean
velocity, and hence the gradient of particle displacement, is large,
which is understandable as strain is a measure of displacement
gradient. One can also notice from
Fig.~\ref{fig:colorCodedQuantities}b the presence of thin shear lines
in the pillars, where particles with large deviatoric strain rate
reside. The width of these shear lines is about twice the diameter of
the particles. These shear lines largely correspond to the moving
boundary between the deformed and undeformed regions in the
pillar. The presence of such shear lines will appear clearer as pillar
size increases, which will be discussed in the later part of the
article.

\begin{figure}[t!]
\begin{centering}
  \includegraphics[width=0.45\textwidth]{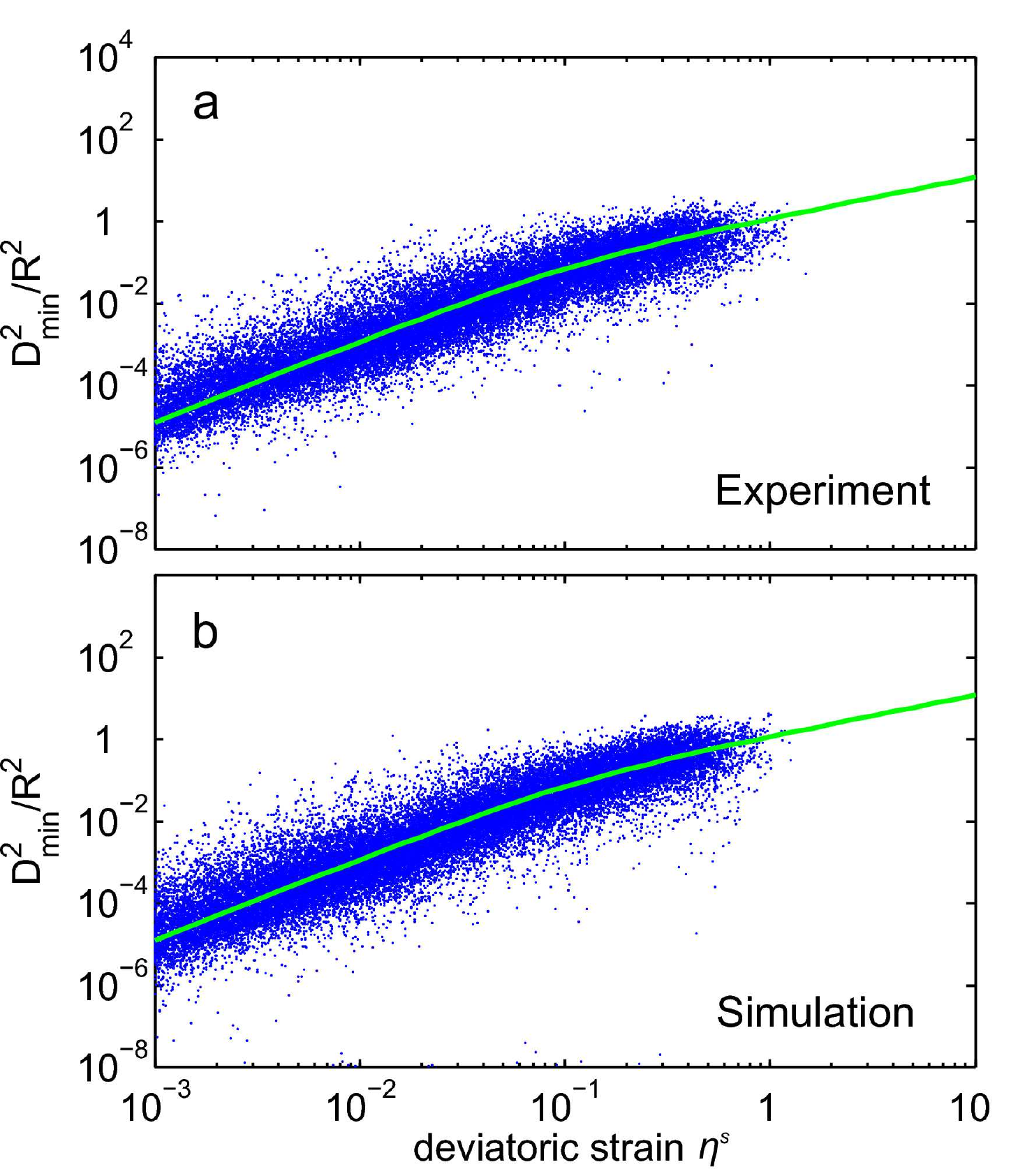}
  \caption{\label{fig:ShearD2minFitting} Fitting of
    $D^2_{\mathrm{min}}/R^2$ with respect to deviatoric strain
    $\eta^s$ for both (a) experiment and (b) simulation.  The data are
    fitted to an exponential crossover equation from quadratic to
    linear scaling (see the main text for details). In (a), the green
    curve is the best-fitting curve for the experimental data. For
    (b), as the fitting results is nearly identical to (a), we plot
    the same fitting curve as in (a) to demonstrate the closeness in
    the fitting result.}
\end{centering}
\end{figure}

Comparing the $D^2_{\mathrm{min}}$ profile in
Fig.~\ref{fig:colorCodedQuantities}c with deviatoric strain rate $J_2$
in Fig.~\ref{fig:colorCodedQuantities}b, it is clear that particles
with larger values of $D^2_{\mathrm{min}}$ are correlated with larger
values of $J_2$, and hence also deviatoric strain $\eta^s$
(Eq.~\ref{eq:J2}). The deviatoric strain $\eta^s$ reflects the local
shear component of affine deformation (shape change), while
$D^2_{\mathrm{min}}$ measures additional particle displacement with
respect to its neighbors that cannot be described by mere shape
change. The positive correlation between $D^2_{\mathrm{min}}$ and
$\eta^s$ is understandable because the larger the value of $\eta^s$
(which usually drives plastic deformation), the error of describing
local particle rearrangement in terms of purely shape change, which is
the definition of $D^2_{\mathrm{min}}$, will more likely to be larger.

\section{Local Deviatoric Strain Driven Particle Diffusion \label{sec:strain_diffusion}}
The positive interdependence between $D^2_{\mathrm{min}}$ and $\eta^s$
motivates us to map out their correlation quantitatively.  Starting
with an initial configuration of the pillar at time $t$ that
corresponds to deformation strain $ \varepsilon = v_ct/H_0$, we fix
the neighbor sampling distance $R_c = 1.25D$ and calculate $\eta^s(t,
\Delta t) $ and $D^2_{\mathrm{min}}(t, \Delta t)$ for each particle in
the pillar using a logarithmic series of time interval $\Delta t \in
[2, 4, 8, ..., 128]/15 $ $R/v_c$. This procedure is then repeated for
at least eight values of initial times $t$ equally spaced by
$\frac{2}{3} R/v_c$. We then plot all the calculated values of
$D^2_{\mathrm{min}}(t,\Delta t)$ with respect to $\eta^s(t, \Delta t)$
on a single plot, using logarithmic axes for both $D^2_{\mathrm{min}}$
and $\eta^s$.  The results of experiment and simulation are shown
together in Fig.~\ref{fig:ShearD2minFitting}. From
Fig.~\ref{fig:ShearD2minFitting}, it can be seen that while for a
given specific value of $\eta^s$, the possible values of
$D^2_{\mathrm{min}}$ are scattered, the existence of statistical
correlation between $D^2_{\mathrm{min}}$ and $\eta^s$ is apparent.  We
find that in the range of small values of $\eta^s$,
$D^2_{\mathrm{min}}$ scales quadratically with $\eta^s$, which
gradually transits to linear scaling at larger values of
$\eta^s$. This is reminiscent of the scaling relationship between the
growth of mean squared displacement (MSD) for a thermally diffusive
particle and time $t$, which is often explained pedagogically by an
unbiased random walker. Indeed, we find that, by considering
$D^2_{\mathrm{min}}$ as MSD, and deviatoric strain $\eta^s$ as time,
the data in Fig.~\ref{fig:ShearD2minFitting} can be fit very well
using the following equation that describes the exponential crossover
of a thermal particle from ballistic to diffusive motion, expected for
a Langevin particle with no memory~\cite{ChaikinLubensky}:
\begin{equation}
D^2_{\mathrm{min}}(\eta^s)/R^2 = 4 \Theta \eta^s - 4\Theta \eta_c^s \left[1-\exp(-\eta^s/\eta_c^s)\right],
\end{equation}
where on the left hand side of the above equation, the calculated
$D^2_{\mathrm{min}}$ is scaled by $R^2$ to render it
dimensionless. $\Theta$ is the dimensionless effective diffusivity
while $\eta_c^s$ takes meaning of ``crossover deviatoric strain''. Our
fitting of the data gives $\Theta = 0.3$, $\eta_c^s = 0.049$ for the
experiment, and $\Theta = 0.3$, $\eta_c^s = 0.05$ for the simulation.

The analogy between $D^2_{\mathrm{min}}$ and MSD, and between $\eta^s
$ and time $t$, may have deep implications.  $D^2_{\mathrm{min}}$
describes the mean-squared non-affine displacement of a particle with
respect to its neighbors and can be naturally identified as an analogy
to MSD. The analogy between deviatoric strain $\eta^s $ and time $t$
implies that, for the granular packings, where there is no thermal
agitation and the system is deformed heterogeneously, the cumulative
deviatoric strain plays the role of time and drives effective particle
diffusion. Argon had originally used bubble raft deformation to
illustrate the concept of shear transformation zone (STZ)
\cite{Argon79Kuo, Argon79}, which emphasizes the affine part of
localized stress-driven processes. Recently, Wang \textit{et al.}
found that cyclic mechanical loading can induce the
nano-crystallization of metallic glasses well below the glass
transition temperature \cite{Wang13}, resulting from stress-driven
accumulation of non-affine displacement of the atoms in the
sample. The concept of shear diffusion transformation zones (SDTZs)
was proposed by the authors to explain the experimental results and to
emphasize the diffusive character of STZs. Our results lend support to
the concept of SDTZ by showing that, even in amorphous granular
packings, where there is no thermal-driven diffusion at all, if the
accumulated local deviatoric strain is large enough (above a few
percent strain), the non-affine displacement of a particle with
respect to its neighbors crosses over to the diffusive limit. This
suggests that SDTZ may be a key concept for a broad range of amorphous
solids.

The analogy between local cumulative shear transformation strain in
athermal amorphous solids and time in thermal systems for particle
diffusion may be rationalized by a ``space-time equivalence" argument,
as follows.  A finite temperature $k_{\rm B}T$ means temporally random
momentum fluctuations, for crystals and non-crystals alike. Even in
crystals, such random momentum fluctuations (due to collision of
multiple phonons) can drive the random walker behavior of a particle,
if these temporal fluctuations can be significant compared to the
potential energy barrier.  But in amorphous solids without spontaneous
temporal fluctuations, there will be nonetheless still another source
of randomness, which is the local spatial structure and structural
response of the amorphous solid. This is indeed what motivated the
``heterogeneously randomized STZ model" \cite{Zhao13, Zhao14}.  In
other words, even if two ``Eshelby inclusions" at different locations
of an amorphous solid transform by exactly the same transformation
strain $\eta$, one reasonably would still expect drastically different
internal particles arrangements and rearrangements inside these zones.
This ultimately is because the local strain $\eta$ is just a
coarse-graining variable, that represents a key aspect of the
structural transformations of a kinetically frozen random cluster, but
not all of its structural information. (This may not be true in simple
crystals, where $\eta$ may entirely capture the entire structure.)
Such structural mutations beyond transformation strain are reflected
in $D^2_{\mathrm{min}}$. The fact that $D^2_{\mathrm{min}}$ will
accumulate linearly with strain at steady state means the structural
mutations from generation to generation \cite{Zhao13, Zhao14} are
largely non-repeating and essentially unpredictable, if starting from
a spatially random configuration at the beginning, even when the
stress condition driving these transformations remains largely the
same.  Our experiment and simulation on compressing amorphous granular
pillars can thus be seen as a ``spatial random number generator" with
the initial configuration as the ``random number seed", in contrast to
more well-known ``temporal random number generator" algorithms; but
both types of algorithms tend to give long-term uncorrelated
increments for the random walker.


\section{Simulation of Size-Dependent Pillar Deformation \label{sec:sizeDependence}}

Having achieved good agreement between experiment and simulation for
the $N =1000$ pillar, we now take advantage of the fact that our
simulation can treat much larger systems than experiment, to study the
size-dependent deformation behavior of the granular pillars by
simulation. Three large-sized pillars, denoted by $N = 4000$, $N =
16000$ and $N = 64000$, are deformed by the top bar moving at the same
deformation speed $v_c$. The aspect ratio of the pillars (2 to 1) is
fixed to be the same value of the $N = 1000$ pillar. As the initial
packing density of the particles in the pillar is also the same, the
initial width of the pillars $W_0$ scales as $\sqrt{N}$.

\begin{figure*}[htb]
\begin{centering}
\includegraphics[width=0.95\textwidth]{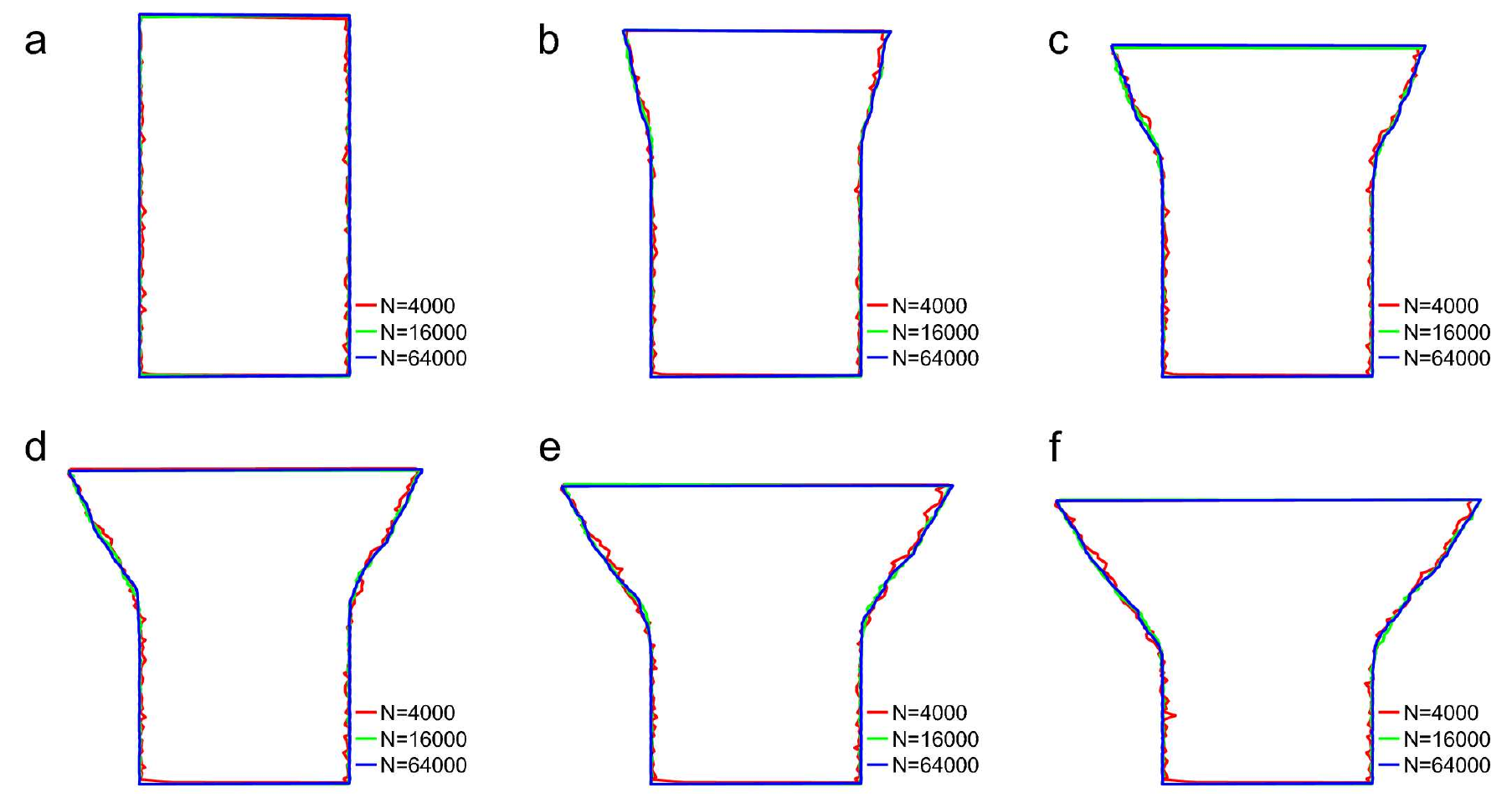}
\caption{\label{fig:pillarboundary} Self-similar evolution of pillar
  shapes during deformation of different-sized pillars. The
  boundaries of three pillars ($N = 4000$, $N = 16000$ and $N =
  64000$) are rescaled and plotted together at the same strain value.}
\end{centering}
\end{figure*}

\begin{figure*}[htb]
\begin{centering}
\includegraphics[width=0.9\textwidth]{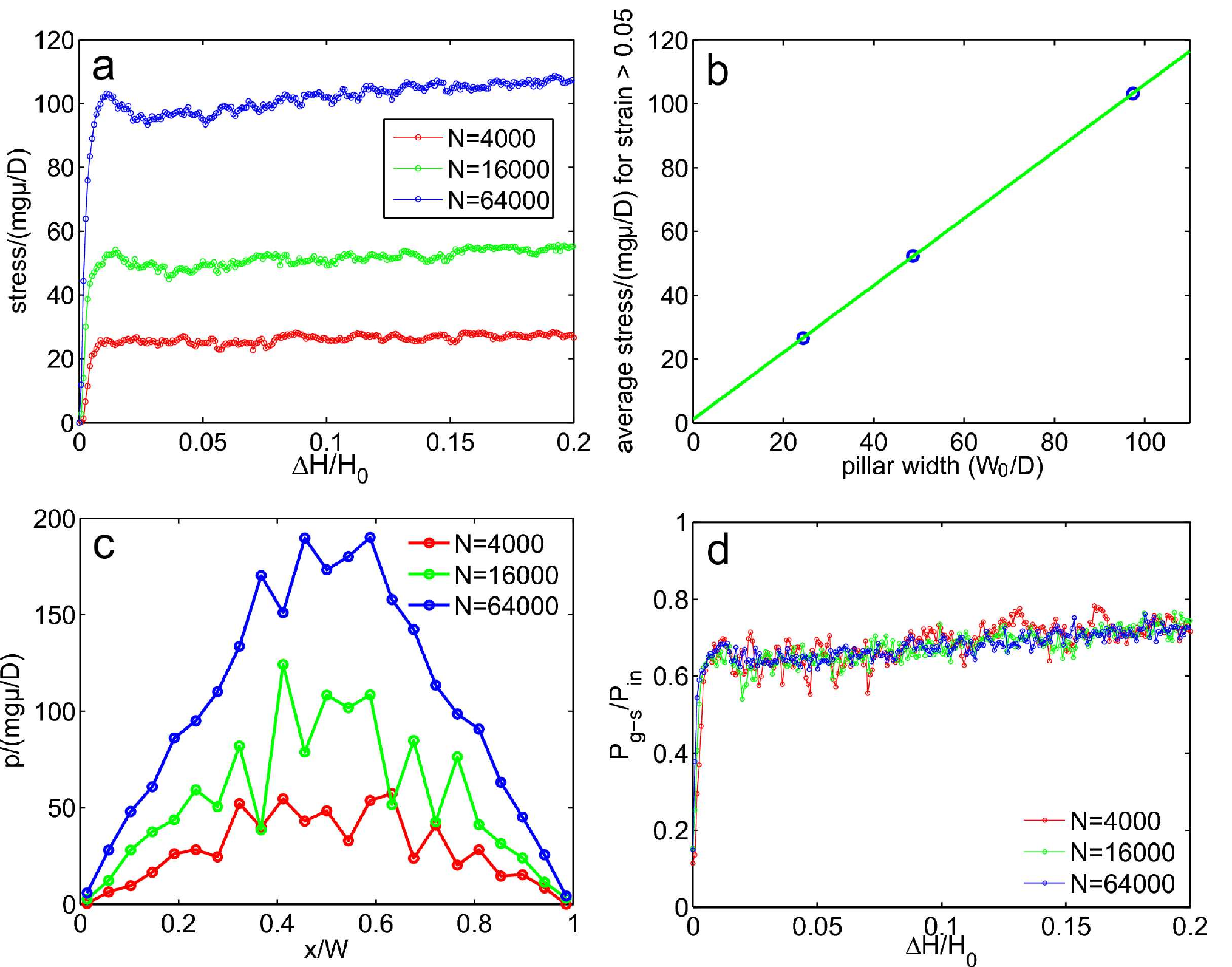}
\caption{\label{fig:stressScaling} Size-dependent flow stress and
  dissipation of input power. (a) Stress strain curves for different
  sized pillars. (b) Linear scaling of average flow stress with
  respect to pillar width $W_0$. The average flow stress is computed
  for the range of strain between 0.05 and 0.2. (c) Local contact
  pressure $p$ between the moving bar and the pillars as a function of
  position $x$ along the contact interface, computed for different
  sized pillars at a representative value of deformation strain in the
  plastic flow regime. The position $x$ is scaled by the width of the
  pillar $W$ at the contact interface. (d) Fraction of input power
  dissipated by the grain-substrate friction as a function of
  deformation strain for different-sized pillars.}
\end{centering}
\end{figure*}


We find the macroscopic shape evolution of the different-sized pillars
is self-similar during deformation.  At the same values of deformation
strain $\varepsilon = \Delta H/H_0$, we extract the boundaries of the
pillars, rescale them by the respective initial pillar width $W_0$,
and plot them together in Fig.~\ref{fig:pillarboundary}. The rescaled
boundaries of the pillars are nearly identical to each other. This
also implies that, the width of the top edge of a pillars $W$ divided
by its original width $W_0$, is to a good approximation only a
function of strain $\varepsilon$ but not the pillar size, namely
$W/W_0 = \chi(\varepsilon)$, where the scaling function $\chi$ does
not depend on the pillar width $W_0$.

We also find that, the average flow stress of the pillars increases
linearly with the initial pillar width $W_0$, as shown in
Fig.~\ref{fig:stressScaling}(a-b). Mathematically, this can be
expressed as $\left<\sigma\right> \propto W_0$, where we define
$\left<\sigma\right>$ to be the average flow stress for strain
$\varepsilon$ between 0.05 and 0.2. This scaling behavior for the flow
stress indicates that, for the 2D disordered granular pillars, the
behavior of ``smaller is weaker'' is exhibited. This is quite
different from the deformation of free-standing metallic glass
pillars, where ``smaller is stronger'' is the general trend
\cite{Wang12Ding, Jang10Greer}.

To understand the surprising size dependence of flow stress, we first
look at the stress distribution in the pillars. In
Fig.~\ref{fig:colorCodedQuantities} we have shown that the grains in
the interior region of the pillar experience larger inter-particle
contact forces, resulting in larger local stress in the interior
region of the pillar. The rate of increase for local stress as a
function of distance to the lateral edges of the pillars is found to
be very close for different-sized pillars.  Such stress non-uniformity
should also be reflected in the local contact pressure between the
moving bar and the pillar. Indeed, we find that the contact pressure
is also spatially rather non-uniform. Fig.~\ref{fig:stressScaling}c
shows that, the local contact pressure increases almost linearly from
near zero at the edge of pillar to saturated values around the center
of contact interface. The maximum values of local contact pressure
scale roughly linearly with pillar width, consistent with the linear
scaling of pillar flow stress.

Since the pillars are deformed quasistatically, most of the
deformation work on the pillars will be dissipated during plastic
flow. The flow stress is therefore closely related to the dissipation
of energy in the systems. We hence study how the energy dissipation in
the pillars changes with pillar size. As the granular particles in the
pillars stand on a substrate, two major mechanisms of energy
dissipation during plastic flow can be identified: one is due to the
grain-substrate friction and the other can be attributed to the
grain-grain friction. The total external power input by the moving bar
into the pillar, denoted by $P_{\mathrm{in}}$, can be calculated as
\begin{equation}
P_{\mathrm{in}} = F_{\mathrm{bar}}v_c = \sigma W v_c.
\end{equation}
We have shown that, compared at the same deformation strain
$\varepsilon$, both the flow stress $\sigma$ and pillar width $W$ are
proportional to the initial pillar width $W_0$.  Hence, the input
power by the external force scales quadratically with $W_0$, namely
$P_{\mathrm{in}} \propto W_0^2$. As most of the input power will be
dissipated in the plastic flow regime, the dissipated power should
also scale with $W_0^2$. To study how the dissipated power is
distributed between the substrate-induced friction and grain-grain
friction, we compute the fraction of input power dissipated by the
grain-substrate frictional force and study its size dependence. The
amount of power dissipated by the grain-substrate friction force,
denoted by $P_{\text{g-s}}$, can be calculated as
\begin{equation}
  P_{\text{g-s}} = \sum_i m_ig\mu v_i,
\label{eq:P_gs}
\end{equation}
where the particle mean velocity $v_i$ has the same definition as in
Eq.~\ref{eq:meanVelocity}, namely the average displacement of the
particle $i$ within a small time interval $\Delta t$. The fraction of
power dissipated by the substrate-induced friction, denoted by
$\kappa$, is then given by $\kappa \equiv
P_{\text{g-s}}/P_{\mathrm{in}}$. We calculate the values of $\kappa$
for different sized pillars and plot them as a function of deformation
strain in Fig.~\ref{fig:stressScaling}d. The result indicates that
$\kappa$ is statistically independent of pillar size. This allows us
to conclude that the amount of input power dissipated by
grain-substrate friction, $P_{\text{g-s}} = \kappa P_{\mathrm{in}}$,
also scales quadratically with pillar size $W_0$, and hence scales
linearly with the number of particles in the pillar $N$. This
effectively means that the number of particles participating in the
plastic flow scales linearly with the total number of particles in the
pillar, which is consistent with the self-similar shape evolution of
the pillars.

The calculated values of $\kappa$ in Fig.~\ref{fig:stressScaling}d
indicate that the majority of deformation work is dissipated by the
friction between the particles in the pillar and the
substrate. Substrate friction therefore must play an important role in
the linear increase of flow stress with respect to pillar width and
the self-similar evolution of pillar shape, which have been shown to
be consistent with each other. The granular pillars in our study are
not truly two-dimensional due to the presence of grain-substrate
friction. This setup is however necessary for stable plastic flow of the
uniaxially deformed granular pillars without cohesive interparticle
interaction. Without the grain-substrate friction, the deformation
behavior of the granular pillars are expected to be quite different,
and the size-dependent deformation behavior observed in this study
(\textit{i.e.} ``smaller is weaker") may no longer hold.

\begin{figure*}[t]
\begin{centering}
\includegraphics[width=1.0\textwidth]{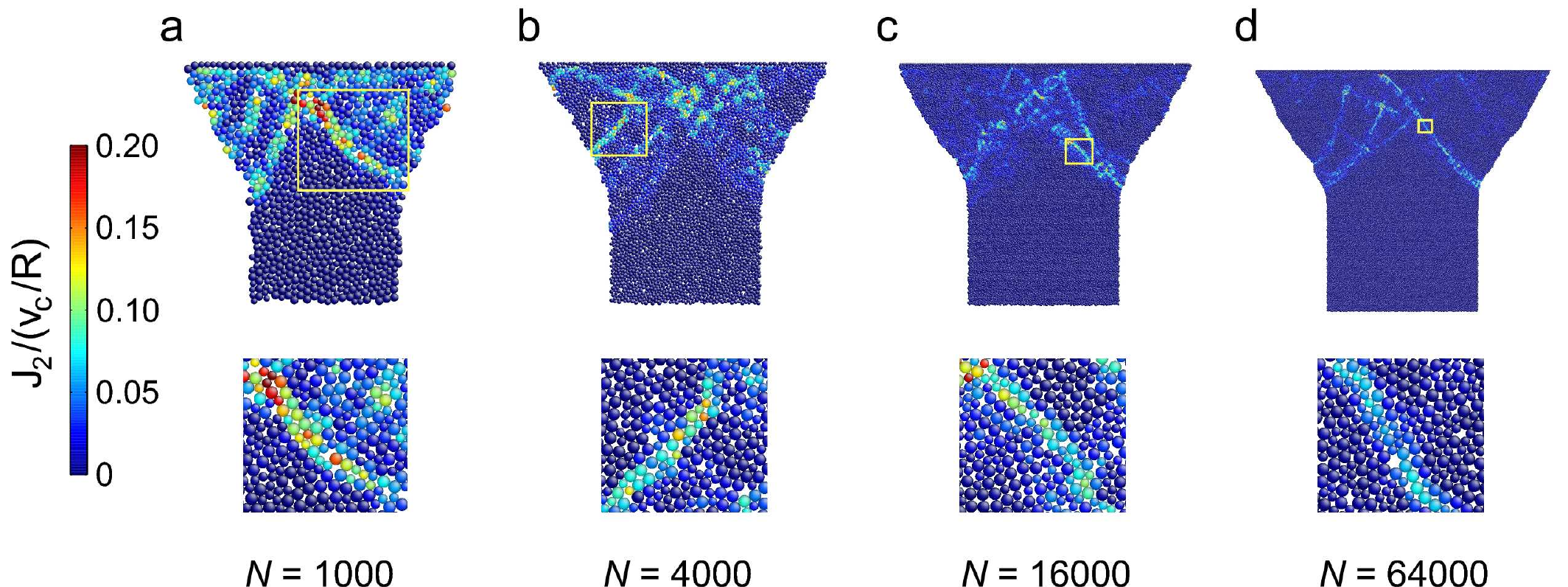}
\caption{\label{fig:shearlines} Deviatoric strain rate $J_2$
  associated with each particle and zoom-in views of the transient
  shear lines in different-sized pillars. Four different-sized pillars
  are compared with each other, which contain 1000, 4000, 16000, 64000
  grains respectively (from left to right). The regions in the pillars
  for zoom-in views are indicated by squares. For each pillar, the two
  configurations of pillars used for $J_2$ calculation are separated
  by time difference $\Delta t = 8/15$ $R/v_c$.}
\end{centering}
\end{figure*}

If the macroscopic shape evolution of the pillars in our systems is
self-similar, then how does the local yielding behavior vary with
pillar size? We characterize the deformation-induced local structural
change of the pillar by computing the deviatoric strain rate $J_2$
associated with each particle between two stages of deformation, using
the same methodology described earlier in the article. We find that,
within a small amount of pillar strain, particles with large values of
$J_2$ organize into thin shear lines, which becomes more evident as
pillar size increases, as shown in Fig.~\ref{fig:shearlines}. These
shear lines orient along the direction about 45 degree to the
direction of uniaxial compression. Clearly, such shear lines form
along the direction of maximum shear stress. The sharpest shear lines
appear predominantly at the moving boundary between the deformed and
undeformed region in the pillars, as we have mentioned earlier when
discussing the combined experimental and simulation study of
small-sized pillars.  A close-up view of these shear lines in
Fig.~\ref{fig:shearlines} indicates that the width of the shear lines
does not change as pillar size increases, maintaining a value about
twice the diameter of a grain. We emphasize that these shear lines are
transient in time. As deformation goes on, new shear lines will form
elsewhere in the pillar, while the particles in the shear lines formed
earlier may not accumulate significant amount of shear strain
continuously. Evidence of such transient shear bands in granular
materials was previously reported in the discrete element simulations
by Aharonov and Sparts \cite{Aharonov02} and Kuhn \cite{Kuhn99,
  Kuhn03}. Maloney and Lema{\^i}tre \cite{Maloney06}, and Tanguy
\textit{et al.} \cite{Tanguy06} observed transient lines of slip in
their athermal, quasistatic simulation of 2D glasses of frictionless
particles, and explained their formation in terms of elastic coupling
and cascading of shear transformation zones. The results of our
combined experiment (Fig.~\ref{fig:colorCodedQuantities}) and
simulation of uniaxial, quasistatic deformation of 2D granular pillars
clearly demonstrate the existence of such transient shear lines, which
carry localized deformation in the granular pillars.

The size-independent width of the transient shear lines is surprising
since the overall macroscopic shape of the pillar is self-similar in
systems of different sizes.  Despite the self-similarity at the
macroscopic scale, the system is not self-similar in how it yields at
the microscopic scale. Since the slip lines are independent of system
size, there must be more of them in larger systems, which is indeed
observed in our simulation. Why the system chooses to be self-similar
at the macroscopic scale but not at the microscopic scale is an
interesting point for future study.



\section{Concluding Remarks \label{sec:conclusion}}

We have carried out combined experiments and simulations of the
quasistatic, uniaxial deformation of 2D amorphous granular pillars on
a substrate. The simulation model developed in this article achieves
excellent quantitative match to the experiment. In particular, we find
that, for the granular packings, the non-affine displacements of the
particles exhibit exponential crossover from ballistic motion to
diffusion-like growth behavior with respect to local deviatoric
strains.  This result is a generalization to inhomogeneous loading of
earlier observations of stress-driven diffusion of particles in
simulated 2D molecular glasses under simple shear or pure shear in the
thermal, quasistatic limit \cite{Ono02,Tanguy06, Lemaitre07,
  Maloney08, Lemaitre09, Martens11}.  Because in our study the ``time"
variable for diffusion, the best-fit deviatoric strain in a
neighborhood, is a local measure of deformation and shear
transformation, we expect that the non-affine displacement should
cross over from ballistic to diffusive behavior in amorphous solids
under any loading conditions.

In metallic glass pillars, the apparent strength of the pillar and
strain localization behavior depends on pillar diameter, manifesting
so-called ``size-dependent plasticity" behavior \cite{Zhao14}. Often,
``smaller is stronger" holds for metallic glasses \cite{Wang12Ding,
  Jang10Greer}. We have shown that for 2D granular pillars on a
substrate, the frictional interaction between the granular particles
and the substrate leads to the opposite size-dependent response,
namely ``smaller is weaker''.

Finally, our combined experiment and simulation study clearly
demonstrate that transient lines of slip form in quasistatically
deformed amorphous granular pillars under uniaxial loading
condition. These system-spanning shear lines carry localized shear
transformations in 2D granular pillars, and their width shows no size
dependence. Altogether, these results could have important
implications for the plasticity and internal structural evolution of
amorphous solids.

\section{Acknowledgement}
We acknowledge support by the UPENN MRSEC, NSF-DMR-1120901.  This work
was partially supported by a grant from the Simons Foundation (\#305547
to Andrea Liu).


%

\end{document}